\documentclass[journal]{IEEEtran}
 \pdfoutput=1 
\ifCLASSINFOpdf
   \usepackage[pdftex]{graphicx}
  \graphicspath{{Figures/}}
\else
\fi
\usepackage{amsmath}
\usepackage[normalem]{ulem}
\usepackage[table]{xcolor}
\definecolor{mygray}{gray}{0.9}
\usepackage{array}
\newcolumntype{C}[1]{>{\centering\let\newline\\\arraybackslash\hspace{0pt}}m{#1}}
\begin{document}
%
\title{Pilot-tone assisted 16-QAM photonic wireless bridge operating at 250 GHz}
%
%
%

\author{Luis Gonzalez-Guerrero,
        Haymen~Shams,~\IEEEmembership{Member,~IEEE,}
        Irshaad~Fatadin,~\IEEEmembership{Senior Member,~IEEE,}
        John~Edward~Wu,
        Martyn~J.~Fice,~\IEEEmembership{Member,~IEEE,}
        Mira~Naftaly
        Alwyn~J.~Seeds~\IEEEmembership{Fellow,~IEEE,}
        and~Cyril~C.~Renaud,~\IEEEmembership{Senior Member,~IEEE}
\thanks{This work was supported by the European Union’s Horizon 2020 research and innovation programme under grant agreement 761579 (TERAPOD). This project has also received funding from the Engineering and Physical Sciences Research Council through the COALESCE (EP/P003990/1) grant.}
\thanks{L. Gonzalez-Guerrero was with the Department of Electronic and Electrical Engineering, University College London, Torrington Place, London, WC1E 7JE, England is now with Universidad Carlos III de Madrid, Departamento de Tecnología Electrónica, 28911 Leganés, Madrid, Spain (e-mail: lgguerre@ing.uc3m.es).}%
\thanks{H. Shams is with CSA Catapult Innovation Centre, Celtic Way, Newport NP10 8BE (e-mail: h.shams@ucl.ac.uk).}
\thanks{J. E. Wu, M. J. Fice, A. J. Seeds and C. C. Renaud are with the Department of Electronic and Electrical Engineering, University College London, Torrington Place, London, WC1E 7JE, England (e-mail: zceejew@ucl.ac.uk; m.fice@ucl.ac.uk; a.seeds@ucl.ac.uk; c.renaud@ucl.ac.uk).}
\thanks{I. Fatadin and M. Naftaly are with the National Physical Laboratory, Teddington, TW11 0LW, U.K. (email: irshaad.fatadin@npl.co.uk; mira.naftaly@npl.co.uk).}}

%
%

\markboth{Journal of \LaTeX\ Class Files}%
{Shell \MakeLowercase{\textit{et al.}}: Bare Demo of IEEEtran.cls for IEEE Journals}
%



\maketitle

\begin{abstract}
A photonic wireless bridge operating at a carrier frequency of 250 GHz is proposed and demonstrated. To mitigate the phase noise of the free-running lasers present in such a link, the tone-assisted carrier recovery is used. Compared to the blind phase noise compensation (PNC) algorithm, this technique exhibited penalties of 0.15 dB and 0.46 dB when used with aggregated Lorentzian linewidths of 28 kHz and 359 kHz, respectively, and 20 GBd 16-quadrature amplitude modulation (QAM) signals. The wireless bridge is also demonstrated in a wavelength division multiplexing (WDM) scenario, where 5 optical channels are generated and sent to the Tx remote antenna unit (RAU). In this configuration, the full band from 224 GHz to 294 GHz is used. Finally, a 50 Gbit/s transmission is achieved with the proposed wireless bridge in single channel configuration. The wireless transmission distance is limited to 10 cm due to the low power emitted by the uni-travelling carrier photodiode used in the experiments. However, link budget calculations based on state-of-the-art THz technology show that distances $>$1000 m can be achieved with this approach.
\end{abstract}


%
\IEEEpeerreviewmaketitle

\section{Introduction}
\begin{figure}[!t]
\centering
\includegraphics[width=8.8cm]{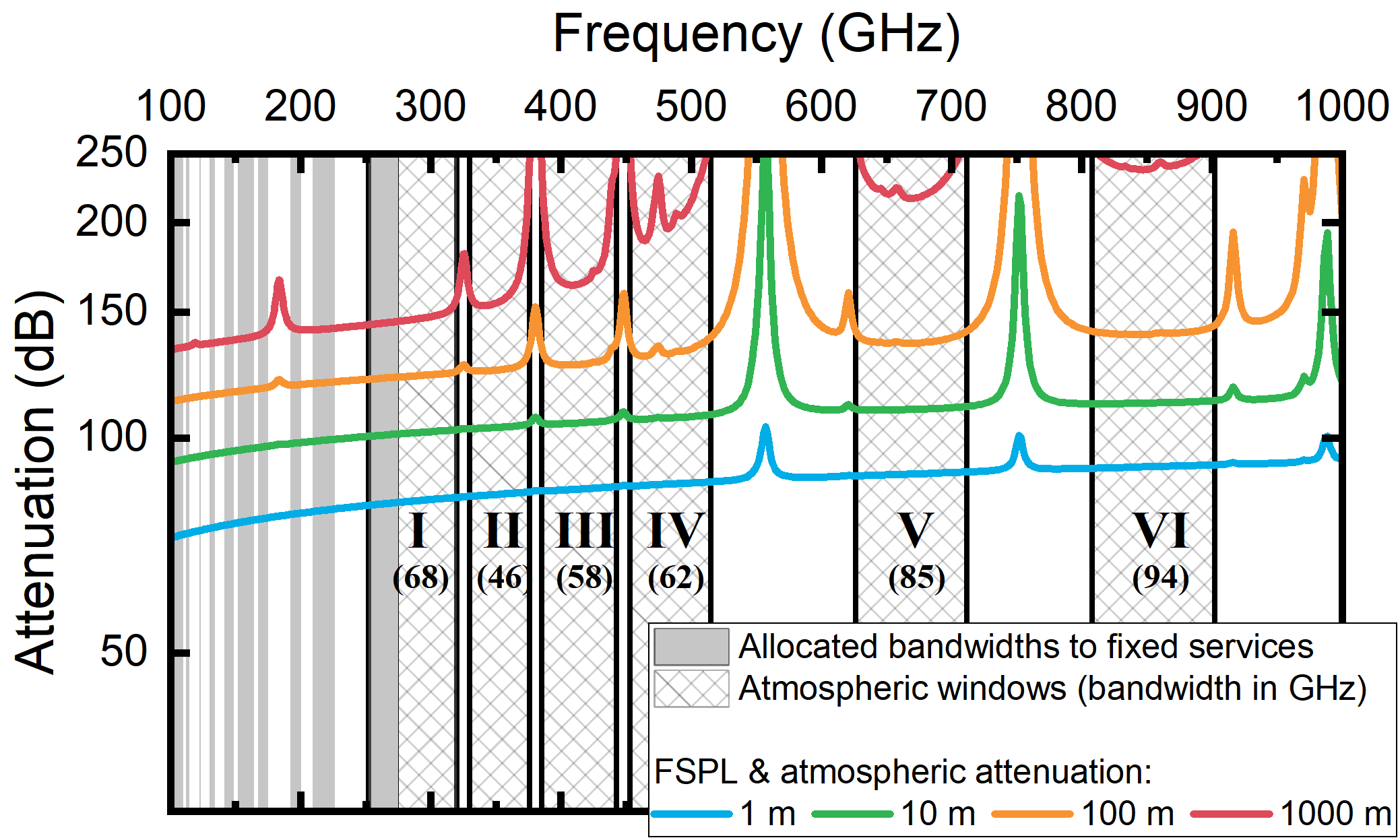}
\caption{Atmospheric attenuation at frequencies from 100 GHz to 1 THz. Values calculated for an atmospheric pressure of 101.300 kPa, temperature of $15^{\circ}$C, and a water vapor density of 7.5 $\text{g/m}^3$}
\label{AtmAtt}
\end{figure}

%
%
%
%
\IEEEPARstart{T}{he} spectrum congestion at radio frequencies (RF) is forcing telecommunication companies to seek a solution at higher frequencies. In their 2017 technology review, Ericsson highlights the importance of the W band (75 GHz -- 110 GHz) and D band (110 GHz -- 170 GHz) for near-future fixed links \cite{Edstam2017}. In that report they also mention the possibility of using even higher frequencies in a more distant future. The reason for this can be seen in Fig.~\ref{AtmAtt}. The windows with relatively low atmospheric attenuation at frequencies above 275 GHz are shown by cross-hatching. Beyond this frequency the spectrum is not currently regulated. Therefore, these windows can be used entirely for data transmission, giving an unprecedented capacity in wireless communications. Included for comparison---in grey---are the windows above 100 GHz currently allocated to fixed services. The widest regulated window in the W and D bands has a bandwidth of 12.5 GHz. As mentioned in the Ericsson report, the ultimate objective is to support data rates of 100 Gbit/s, which may be difficult with such a limited bandwidth. To achieve this capacity in an efficient way, the use of windows beyond 275 GHz will, therefore, be required. Of particular interest for long- and medium-range fixed applications (distances longer than 100 m) is the first unregulated window, which stretches from 275 GHz to approximately 320 GHz. By adding to this the 252 GHz -- 275 GHz frequency range, which is already allocated to fixed services, a continuous bandwidth of 68 GHz is obtained: this is more than 5 times what is currently available in the W and D bands. In the research community, communications using these frequencies are commonly referred to as sub-THz wireless communications.

Among applications envisaged for sub-THz communications, wireless bridges have attracted significant interest \cite{Koenig2014}, \cite{Salamin2018a}. Wireless bridges refer to wireless links connecting two portions of a fibered network. Unlike in wireless fronthaul or backhaul links, in wireless bridges the signal is not demodulated in the receiver antenna unit, but rather up-converted to the optical domain and transmitted to an optical receiver through a second portion of fiber. This application is unique to this part of the spectrum in the sense that it is the only region with potential to achieve wireless data rates comparable to those found in optical networks or, at least, the access and metro parts of such networks. The IEEE standard on high data rate wireless networks \cite{IEEEComputerSociety2017} currently specifies a data rate of 100 Gbit/s. However, if wireless bridges are to be used to connect metropolitan and/or regional parts of the optical network, they will need to support data rates of 200 Gbit/s or more \cite{Heidi2018}. Assuming required coverage distances of $ \ge 100$ m (i.e., transmission limited to individual windows due to high atmospheric attenuation), this scenario calls for high spectral efficiencies (i.e., $>$ 2 bit/s/Hz). 

Apart from these requirements, an easy integration of the wireless transmitter with the optical fiber network is an important requirement for wireless bridges. In this regard, photonic generation of THz signals offers a significant advantage over electronic generation as it directly maps the optical signal to the THz domain \cite{Seeds2015h}, \cite{Nagatsuma2013}. Electronic approaches, on the other hand, need to downconvert the optical signal before THz generation.

In this paper we demonstrate a photonic wireless bridge operating at 250 GHz and using a carrier recovery scheme based on the transmission of a reference pilot tone. Compared to our previous works in \cite{Gonzalez-Guerrero2018} and \cite{Gonzalez-Guerrero2018a} there are two fundamental differences: the first one is that envelope detection is no longer used for demodulation, and the second that the transmission reported here incorporates the second portion of optical fiber (after THz reception) that characterizes wireless bridges.

The rest of the paper is organized as follows. 
Section~\ref{sec.2} provides a review of the various configurations that a photonic wireless bridge can adopt together with notable transmission results published in the literature. 
In section~\ref{sec.3}, the digital signal processing (DSP) employed throughout this paper and the experimental arrangement used for transmission are explained. 
Section~\ref{sec.4} presents system characterization results and briefly discusses frequency and polarization stability considerations. 
Section~\ref{sec.5} discusses the results obtained for a single-channel THz bridge operating at 250 GHz and transmitting 5 GBd 16-quadrature amplitude modulation (QAM) signals (i.e., gross data rate of 20 Gbit/s). In this section, the results for the proposed carrier recovery technique and two different aggregated linewidths are discussed and compared with those obtained with a state-of-the-art phase noise compensation algorithm. 
In section~\ref{sec.6}, the proposed wireless bridge is demonstrated in a wavelength division multiplexing (WDM) network scenario. In this configuration, the full spectrum from 224 GHz to 294 GHz is used for wireless transmission. 
In section~\ref{sec.7}, the experimental results of a 50 Gbit/s wireless bridge are presented. This section also discusses the experimental issues currently limiting the speed and data rate of the proposed system.
Finally, in section~\ref{sec.8}, the manuscript is summarized highlighting its main contributions.

\section{THz wireless bridges}
\label{sec.2}
\begin{figure}[!t]
\centering
\includegraphics[width=8.8cm]{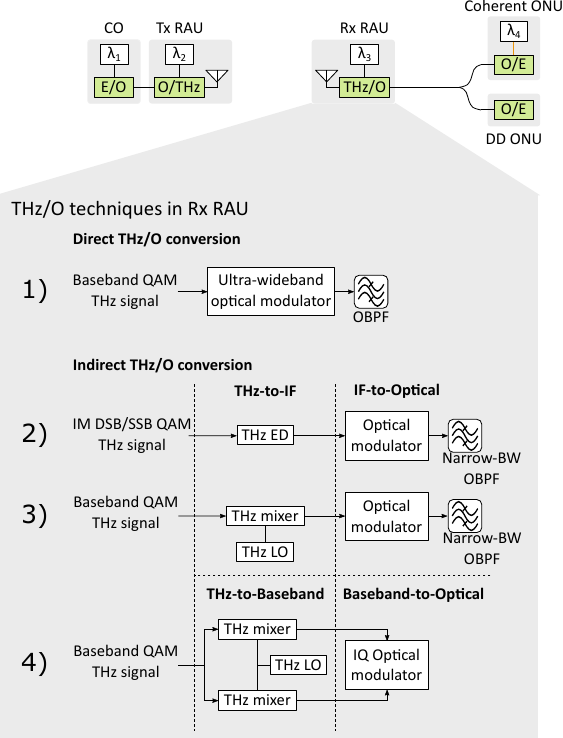}
\caption{Schematic representation of a wireless bridge based on photonic THz generation and methods for THz-to-optical conversion in the Rx RAU (only schemes supporting higher-order modulation are considered). Note that only schemes 2 and 3 are compatible with a direct detection (DD) optical network unit (ONU). For schemes 1 and 4, a coherent optical receiver must be employed to recover the signal. CO: central office; OBPF: optical band-pass filter; IM: intensity modulated; SSB: single sideband; QAM: quadrature amplitude modulation; ED: envelope detector.}
\label{WBconfig}
\end{figure}

Fig.~\ref{WBconfig} (a) presents a depiction of a THz wireless bridge based on photonic THz generation. Fig.~\ref{WBconfig} (b) presents the various methods for THz-to-optical conversion (note that only schemes supporting higher-order modulation are considered) that can be used at the receiver remote antenna unit (Rx RAU). The simplest one is the direct mapping of the incoming THz field to the optical domain via an ultrawide bandwidth optical modulator. Recently, several demonstrations of this concept using plasmonic modulators (which can exhibit remarkably high electrical bandwidths \cite{Burla2019}) have been reported \cite{Salamin2018a}, \cite{Ummethala2018a}. After optical mapping, an optical filter is needed to select one of the data-carrying sidebands. However, since the wavelength separation between carrier and sideband will be at least a couple of nm if transmission is at THz frequencies, a narrow-bandwidth optical band pass filter (OBPF) is not required.

The alternative approach is to down-convert the THz signal first and then perform the up-conversion to the optical domain via a conventional optical modulator. For THz downconversion either homodyne \cite{Koenig2013a}, heterodyne \cite{Wang2018a}, \cite{Kanno2016a}, or direct \cite{Hermelo2017} \cite{Stohr2016} detection can be used. The advantage of the homodyne receiver is that no redundant signal is generated and, hence, it does not require a narrow-bandwidth OBPF to remove it. On the other hand, down-conversion to an intermediate frequency (IF), either with an envelope detector (ED) or heterodyne receiver, simplifies the number of high-frequency components in the receiver RAU, making it a more cost-effective solution.

\begin{table*}[t]
  \renewcommand{\arraystretch}{1.5} 
  \centering
  \caption{Wireless bridges supporting higher-order modulation}
  \label{tab:1}
  \begin{tabular}
  {C{0.9cm} C{2.2cm} C{1.2cm} C{1.2cm} C{1.2cm} C{1.3cm} C{1.1cm} C{0.8cm} C{0.9cm} C{1.4cm}}
  \hline
  \rowcolor{mygray}
  Freq. (GHz) & Rate (Gbit/s); Format & Spectral efficiency (b/s/Hz)&Wireless distance (cm) & Emitter & THz amplifier & Optical distance (km) & THz/O tech. & ONU & Reference\\ 
  \hline
  \hline
  60 & 10/20; QPSK & 1.9 &500/100 & Photonic & Tx & - & 1 & Coh. & \cite{Salamin2018a}\\
  95 & 80; PM-16QAM & 4 & 100 & Photonic & Tx \& Rx & 100 & 3 & Coh. & \cite{Li2014a}\\
  220 & 9; OFDM-QPSK & 0.9 & 50 & Electronic & Tx \& Rx & 40 & 3 & DD & \cite{Koenig2014}\\
  \rowcolor{mygray} 
  250 & 50; 16-QAM & 3.55 & 10 & Photonic & - & 50 & 3 & Coh. & \textbf{This paper}\\
  288.5 & 36; QPSK & 1.81 & 1600 & Photonic & Rx & - & 1 & Coh. & \cite{Ummethala2018a}\\
  300 & 40; QPSK & 1 & 10 & Photonic & Rx & 10.5 & 3 & Coh. & \cite{Kanno2016a}\\
  450 & 13; QPSK & 1 & 380 & Photonic & - & 12.2 & 3 & DD & \cite{Wang2018a}\\ \hline
  \end{tabular}
\end{table*}

Table~\ref{tab:1} shows relevant experiments on wireless bridges supporting higher order modulation formats. For the sake of comparison only transmissions achieving bit error rates (BERs) below the hard-decision forward error correction (HD-FEC) limit of $3.8\times10^{-3}$ are shown. As can be seen, the current highest data rate is 80 Gbit/s---achieved using polarization multiplexing (PM) and 16-QAM modulation. However, this was obtained in a system operating at a carrier frequency below 100 GHz \cite{Li2014a}. Above 100 GHz, systems supporting only quadrature phase shift keying (QPSK)---in either single carrier or multi-carrier (i.e., orthogonal frequency division multiplexing, OFDM) modulation---have been demonstrated so far. 

With the system we propose here, a maximum data rate of 50 Gbit/s was transmitted at a frequency of 250 GHz using 16-QAM signals, achieving a gross spectral efficiency of 3.55 bit/s/Hz. In this proof-of-concept demonstration, the wireless distance was limited to 10 cm due to the limited output power from the photodiode (PD) used to generate the wireless signal. However, as discussed in section~\ref{sec.7}, this distance could be easily extended by using state-of-the-art PDs together with high-gain antennas or combining several PDs in an antenna array.

A particular feature of THz wireless bridges is the number of domain conversions (i.e., E/O, O/THz, etc.) that take place throughout the link. In each conversion process, the phase noise of the LO is added to that of the signal. This can result in a very high level of phase noise at the ONU. To mitigate this, one can base the wireless bridge on direct detection and single-sideband with carrier (SSB-C) signalling. With this approach, a THz ED and a DD optical receiver would be used for THz/O (configuration number 2 in Fig.~\ref{WBconfig}) and E/O conversion at the Rx RAU and ONU, respectively. In this case, only the phase noise of the electrical source driving the optical modulator in the CO would be present at the ONU (assuming optical dispersion-induced phase noise is negligible). However, as noted in \cite{Nagatsuma2013}, the lower sensitivity of THz EDs compared to that of heterodyne mixers, may hinder their implementation in medium- and long-range applications. 

Alternatively, one can use an optical frequency comb generator (OFCG) to produce two phase-correlated tones at the CO \cite{Shams2015}. While this approach has the potential of generating a very pure THz tone \cite{Nagatsuma2016} and removes the need for an optical LO at the Tx RAU, the complexity associated with the OFCG and subsequent demultiplexer might make this approach less attractive than using two-free running lasers. Also, having the optical LO at the Tx RAU has some advantages: first, it enables duplex communication because the Tx RAU laser can also be used to drive the upstream modulator.  Second, the optical LO does not get attenuated by fibre transmission. These two advantages are the same advantages associated with coherent detection in optical networks.

To enable the use of low-cost foundry-fabricated lasers and spectrally-efficient modulation formats without incurring a high penalty, a robust phase noise compensation technique is, therefore, essential in this type of link. In the system we propose here, the tone-assisted technique \cite{Jansen2008} is used for carrier recovery. In this technique, a pilot tone is transmitted together with the data-carrying signal to track the phase noise accumulated throughout the wireless bridge. In the ONU, after analog-to-digital conversion (ADC), this tone is then used to coherently down-convert the signal.

\section{DSP and experimental arrangement}
\label{sec.3}
\begin{figure}[!t]
\centering
\includegraphics[width=8.8cm]{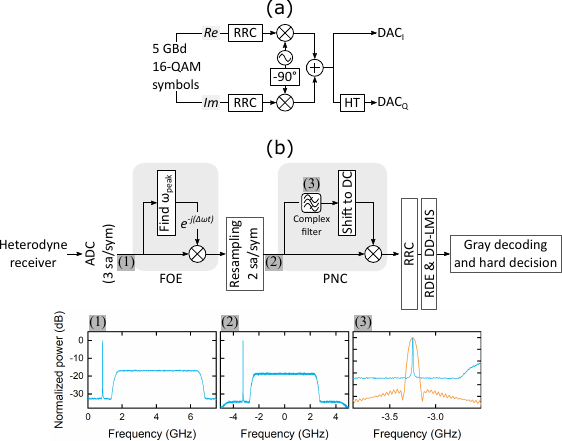}
\caption{(a) Tx DSP for the generation of SSB-SC signals and (b) Rx DSP including the tone-assisted technique for carrier recovery. Insets: (1) received SSB signal sampled at 3 samp/symb; (2) signal after FOE and resampling; (3) Gaussian filter for PNC.}
\label{DSP}
\end{figure}

In this paper, the pilot tone and the signal were generated via the SSB-C modulation format, which was implemented with an IQ modulator and a digital Hilbert transform as shown in Fig.~\ref{DSP} (a). To limit the bandwidth of the signal a pair of root raised cosine (RRC) filters with $ \alpha = 0.1$ were applied before digital up-conversion.

In Fig.~\ref{DSP} (b), the DSP used for tone-assisted carrier recovery is illustrated. The frequency offset estimation (FOE) block comprises a peak search and a down-conversion operation. After resampling to 2 samples per symbol the phase noise compensation (PNC) takes place. This is done by: a) filtering out the pilot tone with a complex filter, b) shifting the filtered tone to DC and 3) mixing the pilot tone with the original unfiltered signal. After PNC, matched filtering is applied and the signal is equalized with the radius directed and decision-directed equalizers (RDE and DDE, respectively).

An extra DSP block with the square QAM-adapted Viterbi Viterbi \cite{Seimetz2008} algorithm was used in this work between the RDE and the DDE. This was only used to compensate a fixed phase offset and a very large averaging block was used. The fixed phase offset results from the biasing points of the I- and Q-components in the optical IQ modulator as detailed in \cite{Gonzalez-Guerrero2018}.  In a practical system, to reduce digital complexity, this offset may be removed by simply adding the proper phase value to the filtered tone \cite{Zhu2018}. Here, since we were constantly changing the modulator biasing points to adjust the pilot-tone to signal power ratio (PTSPR, see 2 paragraphs below), the Viterbi Viterbi algorithm was used to avoid measuring the phase offset at each PTSPR value.

\begin{figure}[!t]
\centering
\includegraphics{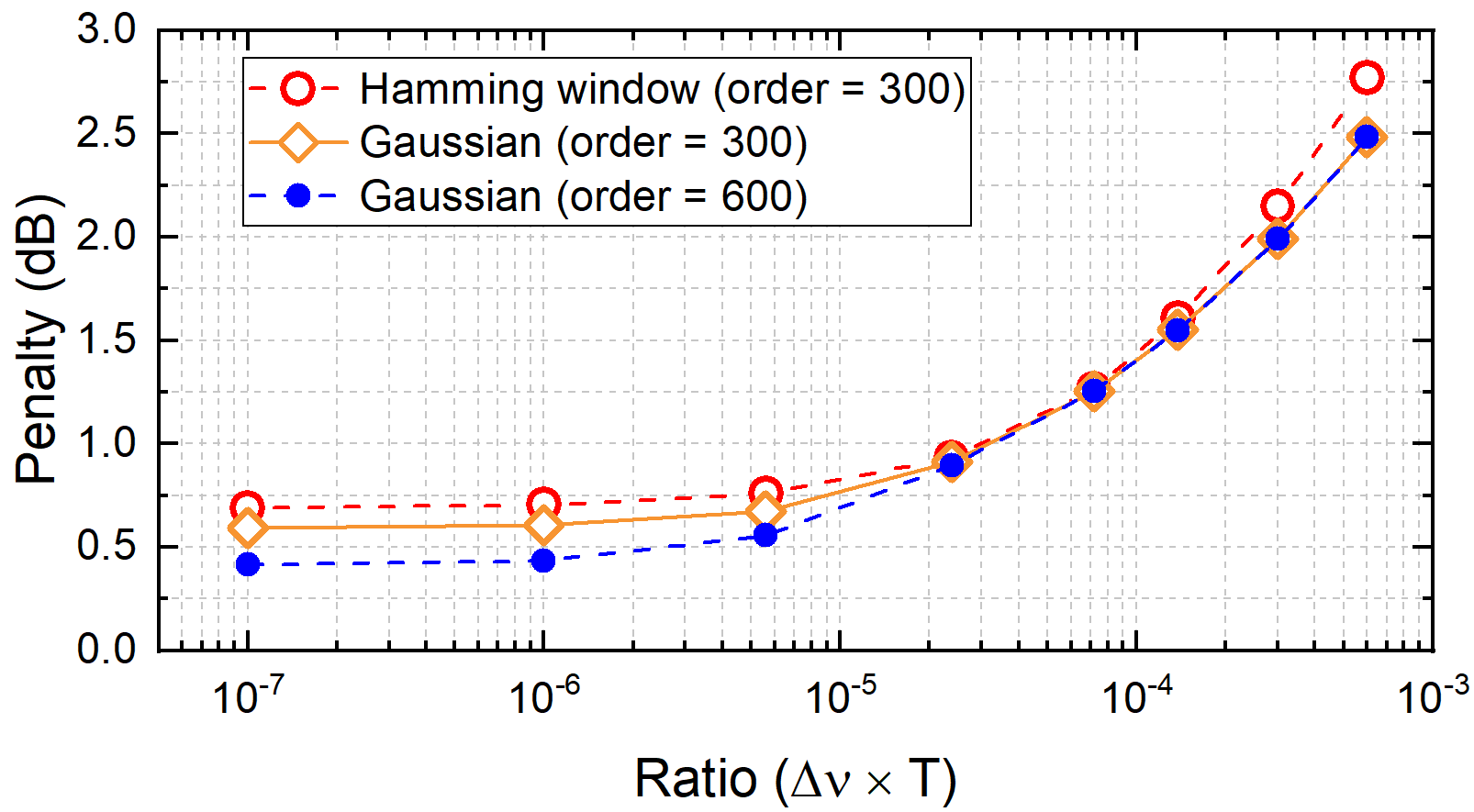}
\caption{Penalty of the pilot tone-assisted technique using a Gaussian and a Hamming-window filter.}
\label{FIRfilters}
\end{figure}

An important design parameter of the pilot-tone assisted carrier recovery is the filter used for PNC. In Fig.~\ref{FIRfilters}, the penalty of two different finite impulse response (FIR) filters is plotted against the linewidth-symbol time $(\Delta\nu\times \textnormal{T})$ ratio (details of the simulations can be found in appendix~\ref{sec.9}). These two filters were chosen due to their previous use in the literature: the Gaussian filter was used in \cite{Morsy-Osman2011} and \cite{Randel2010a}, whereas the Hamming filter was employed in \cite{Zhang2012a} and \cite{Zhang2012}. As can be seen, the Gaussian filter outperforms the Hamming-windowed filter at low and high $(\Delta\nu\times \textnormal{T})$ ratios. The advantage of using a large filter order can also be seen at low ratios, where the required filter bandwidth is small (see section~\ref{sec.5} for guidelines in the optimization of the filter bandwidth with respect to laser linewidth) and the width of the impulse response becomes very large.

\begin{figure*}[!t]
\centering
\includegraphics{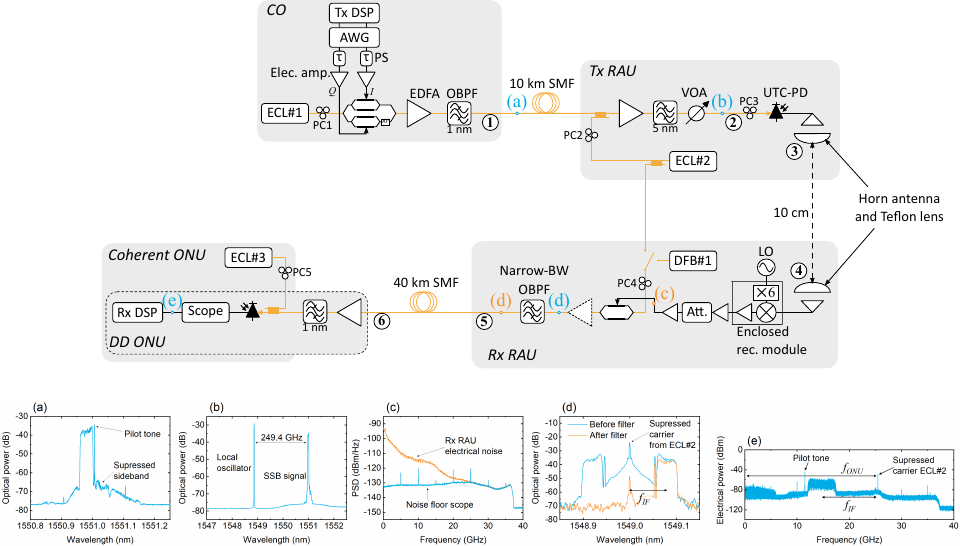}
\caption{Complete experimental arrangement. The numbers throughout the link denote the points at which the signal-to-noise ratio (SNR) was measured (see Fig.~\ref{FreqStab} (b)). The dashed amplifier at the Rx RAU was only used with the coherent ONU. For the DD ONU, this amplifier was placed at the receiver unit. Insets: (a) optical spectrum of the SSB signal generated at the CO; (b) spectrum at the input of the UTC-PD; (c) electrical noise at the Rx RAU (this was measured connecting the output of the 3rd IF amplifier directly to the real-time scope when no THz signal was transmitted); (d) optical spectrum of the signal generated in the Rx RAU before (blue trace) and after (orange trace) the OBPF; and (e) electrical spectrum of the signal received at the ONU. Note that, due to the use of a single-ended PD at the ONU, a rather high down-conversion frequency was used at this unit to avoid the signal-signal beating interference (SSBI). In a practical system, either a balanced PD or a suitable DSP algorithm \cite{Li2017} could be used to mitigate the SSBI and reduce the down-conversion frequency and the required sampling rate of the oscilloscope.}
\label{ExpArr}
\end{figure*}

The complete experimental arrangement used in the transmissions is shown in Fig.~\ref{ExpArr}. For digital signal generation, four $2^{11}$ de Bruijn bit sequences were mapped into the 5 GBd 16-QAM symbols using Matlab. For SSB-C signal generation, the signal was first up-converted to a frequency of 3.25 GHz, giving a guard band (GB) between tone and signal of 500 MHz and a total passband bandwidth of 6 GHz. The resultant waveforms after applying the transmitter DSP were uploaded to an arbitrary waveform generator (AWG) operating at 50 GSa/s and with an analog bandwidth of 12 GHz. The two signals generated in the AWG were time-aligned with two phase shifters and electronically amplified before being fed to an optical IQ modulator. 

The pilot tone-to-signal power ratio (PTSPR), whose optimum value depends on the amount of noise present in the system (see section~\ref{sec.5}), was set by adjusting the DC value applied to the I and Q electrodes in the optical modulator while keeping the phase electrode at the quadrature point. For low PTSPRs, these electrodes were biased close to null; to increase the PTSPRs, the biasing points were progressively moved towards the quadrature point.

At the CO, an external cavity laser (ECL\#1) emitting at a wavelength of 1551 nm was used for data modulation. A single mode fiber (SMF) with a length of 10 km was used to connect the CO to the Tx RAU. The optical signal received at this unit was combined with an optical tone with a wavelength of 1549 nm (giving a frequency separation between the two optical signals of around 250 GHz). To test the system under different amounts of phase noise this tone was generated by either another ECL (ECL\#2 in Fig.~\ref{ExpArr}) or a distributed feedback laser (DFB\#1) with respective linewidths of 28 and 359 kHz.  After optical amplification and filtering, the two optical tones were fed into an unpackaged uni-travelling carrier (UTC) PD by means of a lensed fiber. The optical bandwidth of the PD is limited at low wavelengths by the OH$^-$ absorption peak of the optical fibre attached to it and, at long wavelengths, by the absorption coefficient of InGaAs. This limits the optical bandwidth of the UTC from approximately 1.4 $\mu$m to 1.7 $\mu$m. The maximum optical power available at the input of the lensed fiber during the experiments was measured to be 16.5 dBm.

Horn antennas with a gain of 20 dBi were used for both transmission and reception. A pair of PTFE lenses (diameter of 5 cm and focal length of 7.5 cm) separated by 10 cm were inserted between the two antennas to increase the collimation of the THz beam. Such lens diameter gives a theoretical maximum gain of around 42 dB---according to $G = (4\pi S)/\lambda^2$ where G is the gain, S is the area of the lens, and $\lambda$ is the wavelength of the electromagnetic wave (corresponding to the diffraction-limited lossless case \cite{Pozar2012})---which results in a total 84 dB gain for the two lens-antenna pairs. Note, however, that, due to the short transmission distance used in the experiment, a link budget calculation using the Friis formula would give unrealistic results \cite{Friis1946}.

On the Rx RAU, the signal was down-converted to an IF with an enclosed receiver module (WR3.4MixAMC from Virginia Diodes) consisting of a $\times6$ multiplier, a Schottky barrier diode (SBD)-based second harmonic mixer (SHM), and an IF amplifier. The LO frequency was set to 21.8 GHz to achieve an IF of approximately 12 GHz, as shown in Fig.~\ref{ExpArr} inset (E). After down-conversion, the IF signal was passed through two additional IF amplifiers with a 16 dB attenuator between them, giving a total IF gain factor of 64 dB. The resultant electrical signal was used to drive an intensity modulator (IM), which, depending on the optical receiver used, was biased either at the null point (for coherent ONU) or close to quadrature (for DD ONU). After optical amplification, the signal from the IM was filtered with a narrowband OBPF to suppress the upper frequency sideband. Finally, after propagation through 40 km of SMF, the signal was detected and digitized in the ONU. This unit consisted of a single-ended PD, an 80 GSa/s real-time oscilloscope (analog bandwidth of 36 GHz) and (a) an ECL (ECL\#3) in the case of coherent reception or (b) an EDFA and a 1 nm OBPF in the case of direct detection. Signals of 10 $\mu$s of duration (giving a total number of bits of around $2\times10^5$) were used for bit error counting.

\section{System characterization}
\label{sec.4}

\begin{figure}[!b]
\centering
\includegraphics{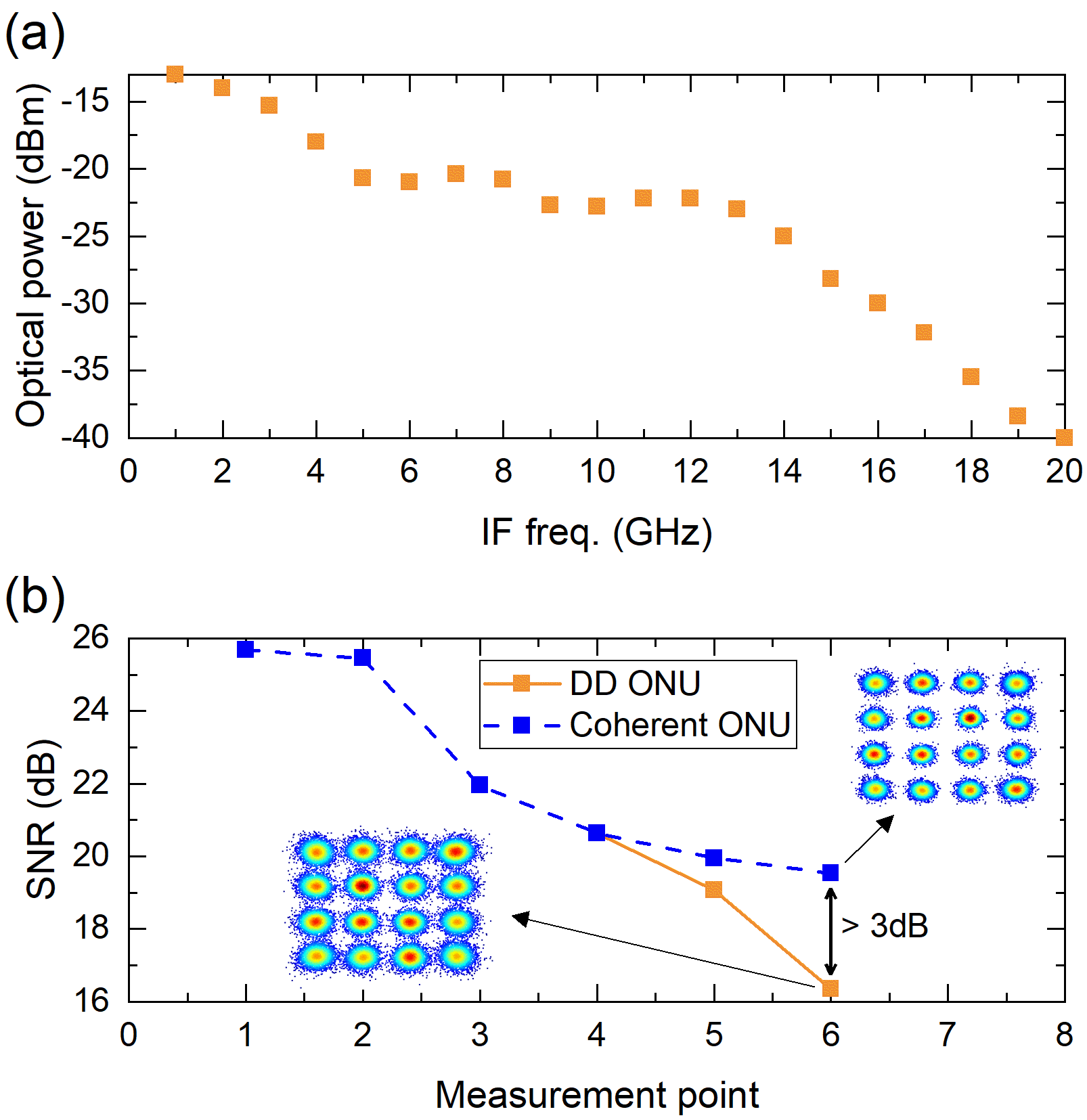}
\caption{(a) IF response of the Rx RAU. For the measurement, the output of the IM (biased at null) was connected to an optical spectrum analyzer, where the power of the low-frequency sideband was measured. The THz frequency was kept fixed and the RF LO was tuned to scan the downconversion frequency. (b) SNR degradation throughout the wireless bridge. The measurements points are highlighted in Fig.~\ref{ExpArr} with circled numbers. For point 3, the transmitter and receiver antennas were placed very close (around 1 cm separation) with no lenses in between. From point 3 onwards, measurements were taken at a constant photocurrent of around 2.15 mA. The constellation diagrams were taken in point 6 (i.e., after the 40-km fibre spool).}
\label{IFandSNR}
\end{figure}

In Fig.~\ref{IFandSNR} (a) the IF response of the Rx RAU (SHM, IF amplifiers and modulator) is shown. As can be seen, the IF response decreases rapidly at low frequencies. This is mainly associated with the frequency response of the $2^{\textnormal{nd}}$ IF amp., which has a strong roll-off close to DC. In spite of this, the overall response still exhibits a sufficiently wide region (bandwidth $>$ 9 GHz) with low power variations (maximum variation of less than 5 dB) to accommodate the 5 GBd signals. The response drop beyond 14 GHz is due to the limited bandwidth of the $3^{\textnormal{rd}}$ IF amp., which has a 3-dB bandwidth specification of 12 GHz. To see the signal degradation throughout the link, SNR measurements were taken at the points specified in Fig.~\ref{ExpArr} with numbered circles. The SNR was calculated as $E[|X|^2]/E[|Y-X|^2],$ where X and Y are the transmitted and received symbols, respectively. As can be seen from Fig.~\ref{IFandSNR} (b), there is a slight drop in SNR (around 1.3 dB) when increasing the wireless distance by inserting the two PTFE lenses. On the other hand, a gain of more than 3 dB is achieved by employing a coherent receiver at the ONU. In the rest of the paper, the coherent ONU is always used.

\begin{figure}[!t]
\centering
\includegraphics{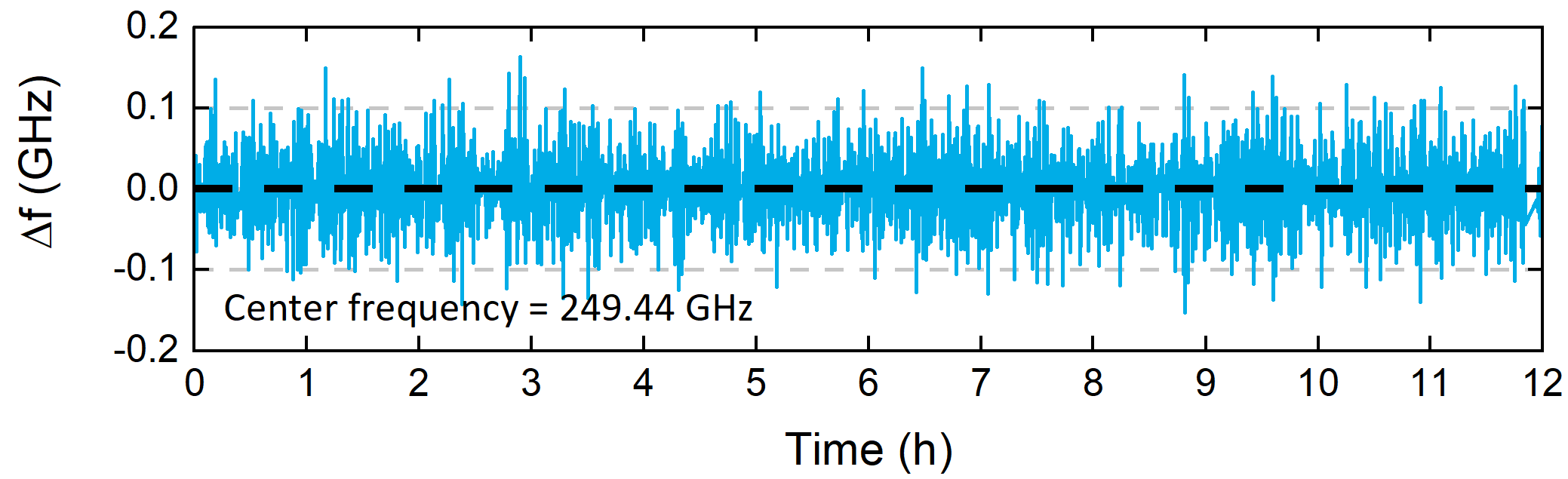}
\caption{Relative frequency fluctuations between the two optical tones used to generate the sub-THz signal.}
\label{FreqStab}
\end{figure}

The frequency stability of the generated signal was also measured by monitoring the wavelength of ECL\#1 and ECL\#2 on a high-resolution optical spectrum analyser (OSA). In Fig.~\ref{FreqStab}, the frequency difference between the two lasers over a period of 12 hours is shown. The frequency separation was found to vary less than $\pm0.2$ GHz across the whole period. This translates into $\pm800$ ppm at a carrier frequency of 250 GHz. While this does not comply with current ITU regulations ($\pm150$ ppm for frequencies between 30 and 275 GHz \cite{ITU1997}), this variation may be reasonable for point-to-point links using large bandwidths, as noted in \cite{Rommel2016}. This stems from the fact that, for the highest range of frequencies currently regulated, ITU specifies a long-term frequency-tolerance objective based on the occupied bandwidth, whereby fluctuations lower than 2\% of the transmission bandwidth would be acceptable. If this criterion is to be applied to high-bandwidth sub-THz communications, then, for a $\pm0.2$ GHz variation, this requirement would be met with a signal bandwidth of 10 GHz. This bandwidth is exceeded in section~\ref{sec.7}, where a 13.75-GHz signal is transmitted through the link. 

Apart from jeopardizing the compliance of ITU spectral mask requirements, these fluctuations can have a strong impact on the long-term performance of the system. These fluctuations arise due to the carrier's low frequency noise and are only observable at long measurement times \cite{Kikuchi2011}. While the duration of the acquired signals (10 $\mu$s) may not be long enough for the low frequency noise to have a noticeable effect, the adoption of the FOE DSP block shown in Fig.~\ref{ExpArr} would ensure that these frequency fluctuations are tracked and compensated for in a real system. This FOE block works by searching for the tone in the frequency domain, so an important parameter is the required frequency resolution (determined by the sampling rate and fast Fourier transform length) to correctly locate the peak. In \cite{Morsy-Osman2011}, it was found that resolutions as low as approximately 14 MHz can be used with no penalty.

Polarization stability is another important aspect to take into consideration. In the current system, up to 5 polarization controllers of manual operation were used throughout the link. In a real system, however, these would not be practical for obvious reasons. To make the system robust against polarization rotations several approaches could be adopted: photonic integration, automatic polarization control, or polarization-diversity techniques. Using a monolithically integrated chip combining the laser and the modulator at both the ONU and Rx RAU would solve the problem in these two units. For the rest of PCs, automatic polarization control using proper feedback electronic circuits and polarization transformers could be adopted.

As noted in \cite{Li2018a}, ideally, the PD used for sub-THz signal generation should be insensitive to polarization (this mainly depends on the way PDs are illuminated---whether they are vertically-illuminated or they use a waveguide). This way, PC3 (as labelled in Fig.~\ref{ExpArr}) would not be necessary. Otherwise, one may use the DC term from the UTC-PD as the feedback signal for automatic control. Regarding PC2, the adoption of a polarization-diversity optical receiver \cite{Glance1987} (in this case split between the Tx and Rx RAU) would eliminate its need. However, an extra pair of antennas, another UTC, and another mixer would be necessary. As this is the same hardware effort that would be required for polarization multiplexing, this approach may not be the most efficient one. Alternatively, one may get rid of PC2 by using a polarization-diversity transmitter at the ONU and proper coding, as reported in \cite{Erklllnc2017b}. The coding technique used in this work---called Alamouti coding---uses two orthogonal polarization sates to send the same information, so that the signal can be recovered regardless of the relative polarization state of the local oscillator.

\section{Phase noise tolerance}
\label{sec.5}

\begin{figure}[!t]
\centering
\includegraphics{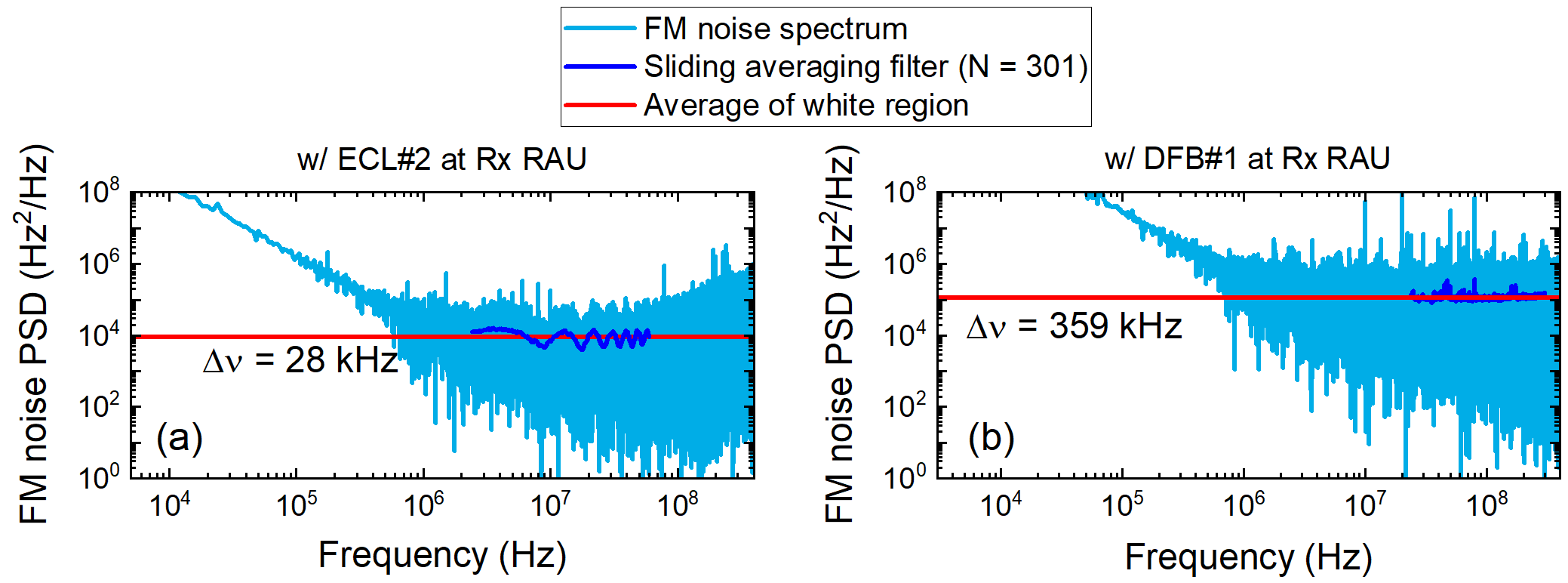}
\caption{FM noise spectrum of the received tone at the ONU when using, at the Tx RAU, (a) ECL\#2 and (b) DFB\#1. The dark-blue curve shows the white region of the FM noise after being passed through a sliding averaging filter with 301 taps. The red curve is the average of the filter output at this region (which is then multiplied by $\pi$ to calculate the Lorentzian linewidth).}
\label{FM noise comp}
\end{figure}

In the first instance, BER measurements were taken with ECL\#2 feeding both the Tx RAU and Rx RAU. Then ECL\#2 was replaced by DFB\#1 in the Rx RAU to see the penalty associated with a high level of phase noise. For both lasers, the frequency modulation (FM) noise spectrum of a 0.5 ms signal received at the ONU (i.e., including all sources of phase noise throughout the link) was measured when no data was being transmitted. The received tone was digitized and processed digitally offline (down-converted, filtered and resampled) before estimating the white frequency noise component and the Lorentzian linewidth \cite{Fatadin2013a}. In Fig~\ref{FM noise comp}, the FM noise spectrum for each arrangement is shown. Linewidths of 28 kHz and 359 kHz were estimated when using ECL\#2 and DFB\#1, respectively, at the Rx RAU. Note that the parameter affecting the performance of a PNC algorithm is the linewidth (which is calculated from the FM white noise) as discussed in \cite{Kikuchi2011}. The distortions caused by the lower-frequency FM noise are taken care by the FOE block. Here, however, we focus the analysis on the PNC algorithm.

To provide a reference for comparison, transmission experiments were also carried out using the blind PNC technique described in \cite{Pfau2009a}. This algorithm has been reported to achieve "nearly optimum linewidth tolerance" \cite{Zhou2014a} and, thus, is a good reference for comparison. The disadvantage of this algorithm, on the other hand, is its high implementation complexity, which arises from the high number of test phases required to produce an accurate estimation. For 16-QAM, for example, 32 test phases are required. A hardware complexity comparison between the two PNC schemes can be found in \cite{Morsy-Osman2011}. When using this algorithm, the IQ optical modulator at the CO was biased at null (since no reference pilot is needed) and a FOE based on the FFT
was employed \cite{Selmi2009}.

\begin{figure}[t]
\centering
\includegraphics{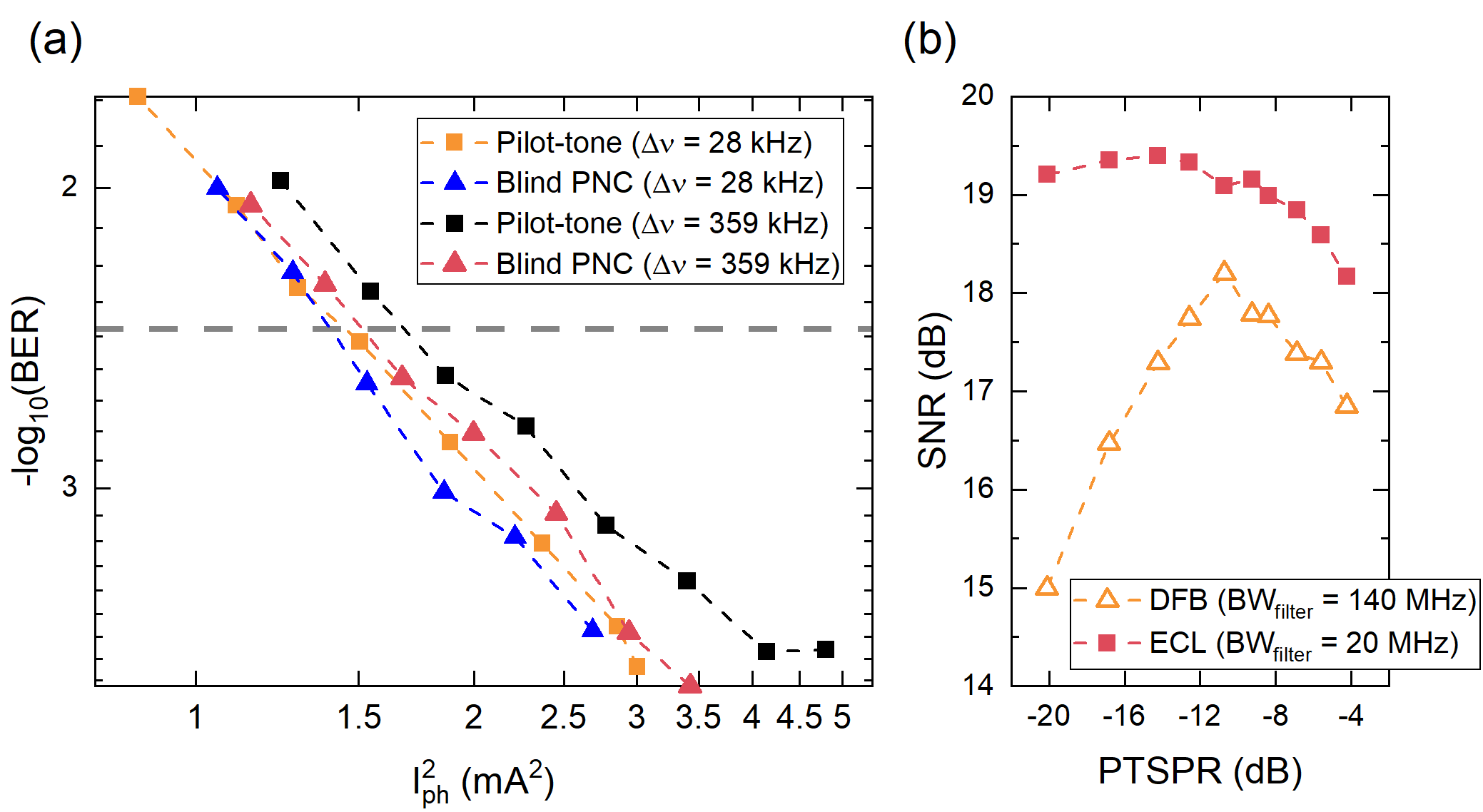}
\caption{(a) BER curves for various transmission arrangements, and (b) SNR against PTSPR for ECL\#1 (squares) and DFB\#1 (triangles)---the optimum filter bandwidth for each case is stated in the legend.}
\label{PNcomparison}
\end{figure}

In Fig.~\ref{PNcomparison} (a), the BER curves for all transmission configurations are shown against the UTC photocurrent squared (proportional to the generated THz power). A penalty of less than 0.65 dB was found when switching from ECL\#2 to DFB\#1 while using the pilot-tone assisted technique. The reason for this penalty is related to the PTSPR, which depends on the amount of FM noise in the system. For low levels, a narrow digital filter can be used to select the tone, minimizing the amount of additive white Gaussian noise (AWGN). This enables the use of low PTSPRs and, hence, biasing the IQ modulator closer to null, where the power---and, thus, the SNR---of the data-carrying sideband is maximized. On the other hand, as the linewidth broadens, a wider filter is needed to correctly track the faster phase distortions. This increases the level of AWGN. To compensate for this and achieve a decent SNR in the filtered tone, the PTSPR must be increased. This, however, compromises the SNR of the signal as less power is allocated to it. Hence, it is necessary to find the right balance (i.e., an optimum PSPR). In Fig.~\ref{PNcomparison} (b), the SNR (for a squared photocurrent of around 4.7 $\textnormal{mA}^2$) versus the PTSPR is plotted for each laser. The optimum PTSPR was found to be -15 dBm and -11 dBm for ECL\#2 and DFB\#1, respectively. This value of PTSPR was then used for all the BER points shown in Fig.~\ref{PNcomparison} (a). The blind PNC, on the other hand, exhibits a penalty of less than 0.35 dB when the Rx RAU linewidth is increased to 359 kHz. Regarding the comparison between the two PNC schemes, the pilot-tone technique exhibits penalties of only 0.15 dB and 0.46 dB when using ECL\#2 and DFB\#1, respectively.

To see the evolution of each technique for different linewidths, Monte-Carlo simulations were carried out (details of the simulations can be found in the appendix), the results of which are plotted in Fig.~\ref{simulations}. In the case of the blind PNC, two different averaging filter orders (2N) were tested: 12 and 26. As expected, a higher order is more efficient at low $(\Delta\nu\times\textnormal{T})$ values, where the phase noise distortion per sample is low and a long averaging block can be used to maximise the SNR. At high $(\Delta\nu\times\textnormal{T})$ ratios, however, the order must be decreased to correctly track the faster phase noise deviations caused by the broadening of the linewidth. The sensitivity decrease of each technique when switching to DFB\#1 agrees well with the simulations. However, in terms of relative performance between the two techniques, the pilot-tone scheme seems to suffer some extra penalty ($<0.15$ dB) in our arrangement. This is likely to be due to the fact that, in the transmission experiments, the PTSPR was only optimized for the highest value of photocurrent and then kept fixed for the whole BER curve. In the simulations, however, the PTSPR was optimized for each value of SNR.

Based on the experimental and simulation results, we believe that the pilot tone-assisted technique offers a good compromise between performance and complexity. Apart from the lower intrinsic complexity of this technique compared to that of the blind PNC \cite{Morsy-Osman2011}, the former does not require differential encoding and decoding---unlike the latter. This adds up to the overall DSP effort of the blind PNC (this is not included in the analysis in \cite{Morsy-Osman2011}). Finally, with the carrier recovery we propose in this paper, the FOE block is readily implemented as it also makes use of the transmitted tone. This is in contrast with the blind PNC technique, which requires a completely different algorithm for FOE.

\begin{figure}[t]
\centering
\includegraphics{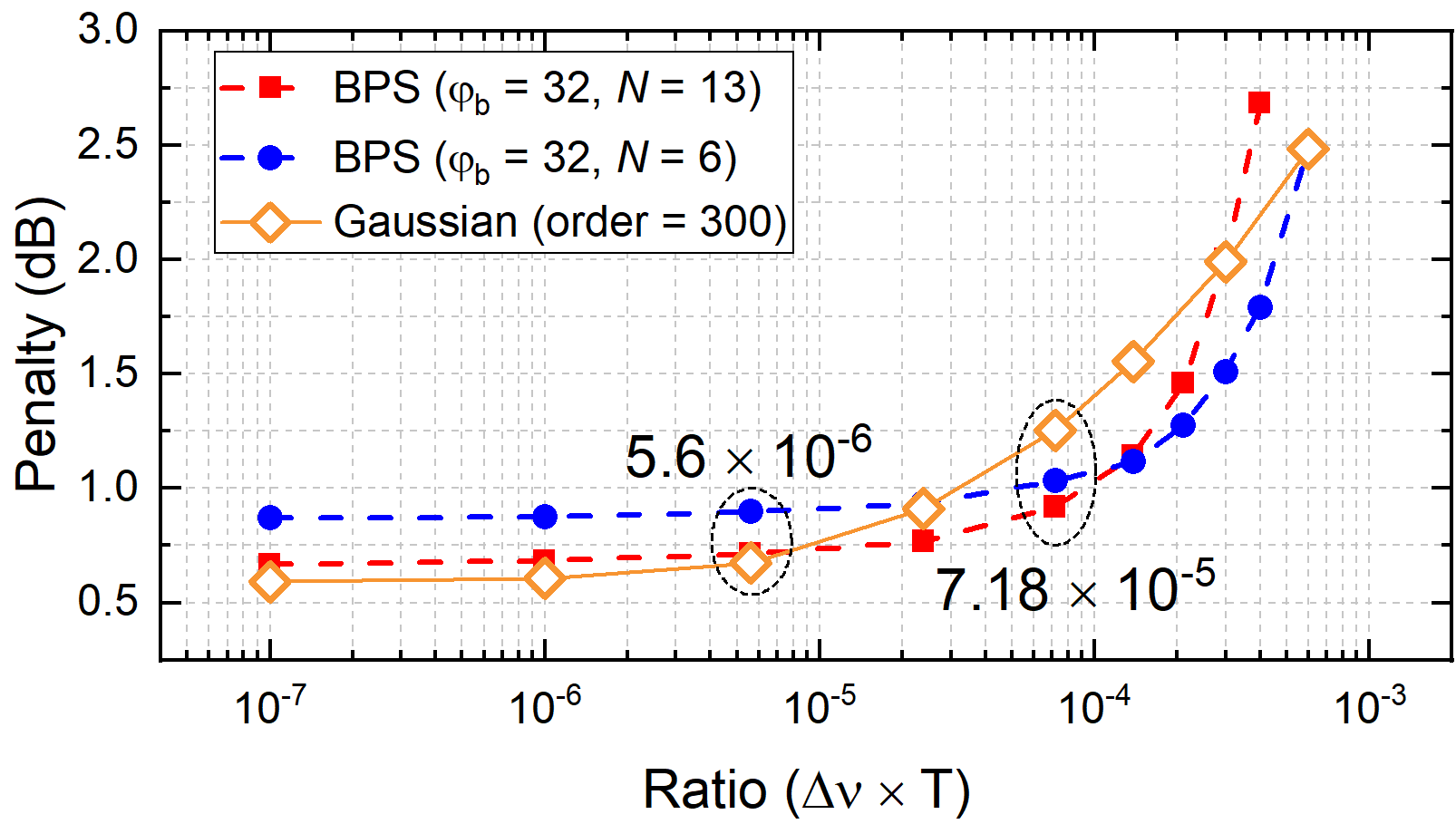}
\caption{Simulation results comparing the blind and pilot tone-assisted PNC techniques. The results obtained at the $(\Delta\nu\times\textnormal{T})$ ratios used in the transmission experiments are marked with circles.}
\label{simulations}
\end{figure}

\section{Wireless bridge on a WDM network}
\label{sec.6}

In reality, the wireless bridge is likely to be part of a
multiuser network where a single CO serves several ONUs
by means of wavelength division multiplexing (WDM). To
demonstrate this type of scenario, a CO supporting five optical channels was built and integrated with the wireless bridge of section~\ref{sec.3}. In the Tx RAU, a narrow OBPF was used to demultiplex one channel at a time. Keeping the wavelength of the Tx-RAU optical LO fixed for all five channels, the full spectrum from 224 GHz to 294 GHz was used in the transmissions.

\begin{figure}[b]
\centering
\includegraphics[width=8.8cm]{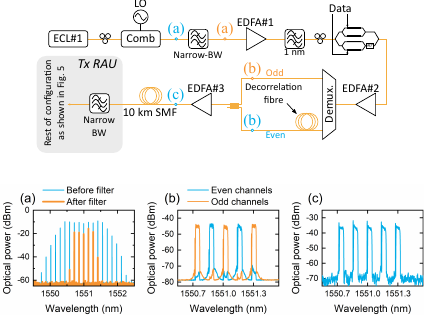}
\caption{Experimental arrangement for generation of multiple optical channels at the CO. The optical comb is generated with a dual-drive Mach-Zehnder modulator (DD-MZM) as detailed in \cite{Sakamoto2007}. Inset (a) shows the comb lines before (blue trace) and after (orange trace) optical filtering. Inset (b) shows the even (blue trace) and odd (orange trace) channels before being recombined (in the moment of capturing the spectrum, the IQ modulator was biased close to null). Inset (c) displays all the channels and their pilot tones after EDFA\#3.}
\label{WDM}
\end{figure}

In Fig.~\ref{WDM}, the experimental arrangement used for the generation of the five optical channels is shown. An optical frequency comb generator (OFCG) based on external modulation and a narrowband OBPF were used to generate the 5 optical lines. An OFCG was employed because not enough ECLs were available at the time of performing the experiment. The maximum line separation supported by the OFCG (around 17.4 GHz---limited by the bandwidth of the electrical amplifier driving the external modulator) was chosen in order to minimize the odd/even channel leakage introduced by the decorrelation stage. This unit was composed of a multiport programmable optical filter and a fiber propagation stage with a physical length mismatch between the odd and even branches of 4 m (corresponding to approximately 90 symbols). Since no such decorrelation unit will be required in a real system, an even smaller channel separation will be possible.

\begin{figure}[t]
\centering
\includegraphics{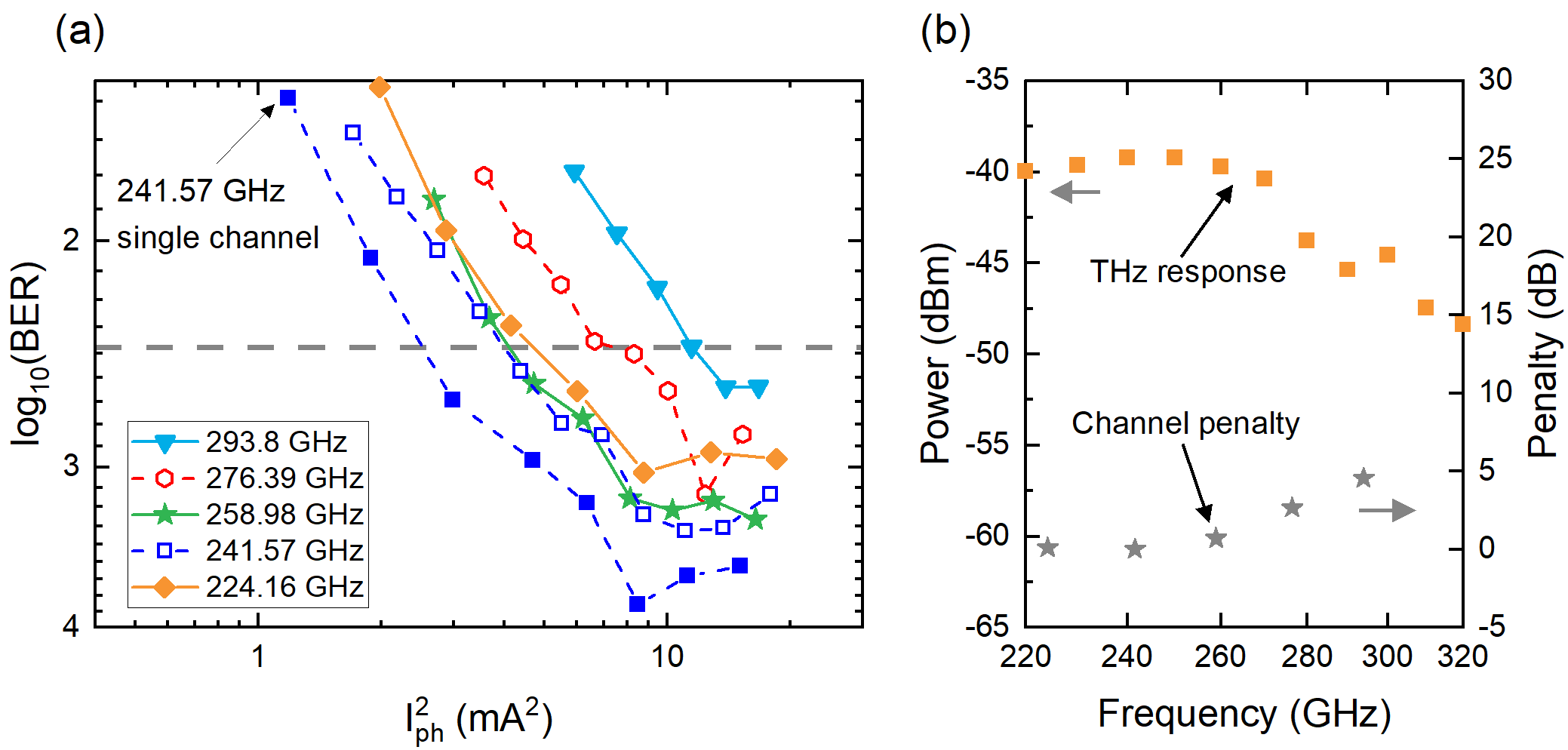}
\caption{(a) BER vs. squared photocurrent for each channel in the multiple-channel configuration and also the 241.57 GHz channel in the single-channel configuration and (b) THz response of the system and penalty of each channel. The response was measured by transmitting an unmodulated tone and measuring the output power from the enclosed receiver with an electrical spectrum analyzer.
The THz frequency was scanned by tuning ECL\#2. The RF LO frequency was also tuned to ensure downconversion to the same IF (10 GHz).}
\label{WDMr}
\end{figure}

In Fig.~\ref{WDMr} (a), the BER curve of each channel is plotted
against the squared photocurrent. To quantify the penalty
associated with the multi-channel operation, the BER curve
of a single-channel transmission was also measured. In this case, the second narrow-bandwidth OBPF in Fig.~\ref{WDM} was placed immediately after EDFA\#3 (i.e., before the 10 km of SMF) so that only one channel propagated through the fiber. EDFA\#3 was operated at maximum output power (14.5 dBm) and the total optical power launched into the fibre (i.e., after filtering) was about 2 dBm. In the case of multi-channel transmission, EDFA\#3 was connected directly to the 10-km fibre spool and optimization of its power output was required to avoid non-linear effects. A power of 7.5 dBm was found to be optimum and a penalty of $<$ 2 dB at the FEC limit was measured between the single- and multiple-channel transmissions. It is important to mention that the power optimization was achieved by adjusting the amplifier's pump current. Less penalty between the single- and multiple-channel configuration may be obtained by operating the EDFA at maximum pump current and inserting a variable attenuator after it \cite{Kikuchi1990}. 

The penalty of each channel with respect to the
241.57 GHz channel (in multiple-channel operation) is plotted in Fig.~\ref{WDMr} (b). The THz response of the system (UTC-PD and SHM) is also shown in this figure. A strong correlation is evident between the two curves. The low power from the UTC used in these experiments (see section~\ref{sec.7}) limited the frequency range of our multi-channel system. As stated in the introduction, only the frequencies above 252 GHz are suitable for fixed communications, rendering the first 2 channels (224.16 and 241.57 GHz) not currently viable for this purpose. However, these results show the large versatility of a wireless bridge based on THz photonic generation and its very easy integration with a WDM network.

In the case of polarization multiplexing (Pol-Mux) WDM networks, the Tx-RAU and Rx-RAU configurations shown in Fig.~\ref{ExpArr} should be extended to support polarization diversity. As shown in \cite{Li2014a}, this would entail the incorporation of another pair of antennas, a UTC, THz mixer, and MZM as well as three polarization beam splitters (PBSs). If Pol-Mux is not required at the second portion of optical fibre (i.e., after the Rx-RAU), one can avoid the Rx-RAU PBS by mapping the signal received by each antenna to opposite optical sidebands  (details of this technique, referred to as twin-SSB, can be found in section~\ref{sec.7}).  This approach, however, requires a complex modulator and takes twice as much optical bandwidth as the previous approach.

\section{50 Gbit/s wireless bridge}
\label{sec.7}

\begin{figure}[t]
\centering
\includegraphics{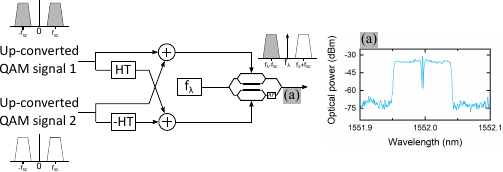}
\caption{DSP employed to generate the twin-SSB signal. Inset (a) shows the optical spectrum of the twin-SSB signal generated after the IQ modulator.}
\label{TwinDSP}
\end{figure}

To investigate the maximum capacity of the system, 40 Gbit/s signals were sent through the link. To achieve this data rate, two types of signals were tested: a 10-GBd SSB signal and a 5-GBd twin-SSB signal \cite{Puerta2017}\cite{Puerta2017a}. The latter was generated by multiplexing two 5-GBd channels in the digital domain using the DSP shown in Fig.~\ref{TwinDSP}. The primary advantage of this technique is that it uses both sidebands to convey information. Compared to the SSB technique, this halves the transmitter bandwidth required to produce a certain data rate . 

On the other hand, in the twin-SSB signal, there is an overlap between the data-carrying signal and the supressed sideband. When this modulation is implemented via an optical IQ modulator, the optical sideband suppression ratio (OSSR) is approximately equal to the extinction ratio (ER) of the IQ modulator \cite{Erkilinc2016}. In the system reported here, the OSSR was around 30 dB (see Fig.~\ref{ExpArr} (a)), which matches the specification value for the ER of our IQ modulator (Covega 086-40-16-SFF). It can be seen from point 1 in Fig.~\ref{IFandSNR} (b), that the SNR of the generated SSB signals is around 26 dB. As this is lower than the measured OSSR value, it is believed that the twin-SSB signals are not limited by OSSR but rather by electrical noise from the AWG and electrical amplifiers.  

\begin{figure}[t]
\centering
\includegraphics{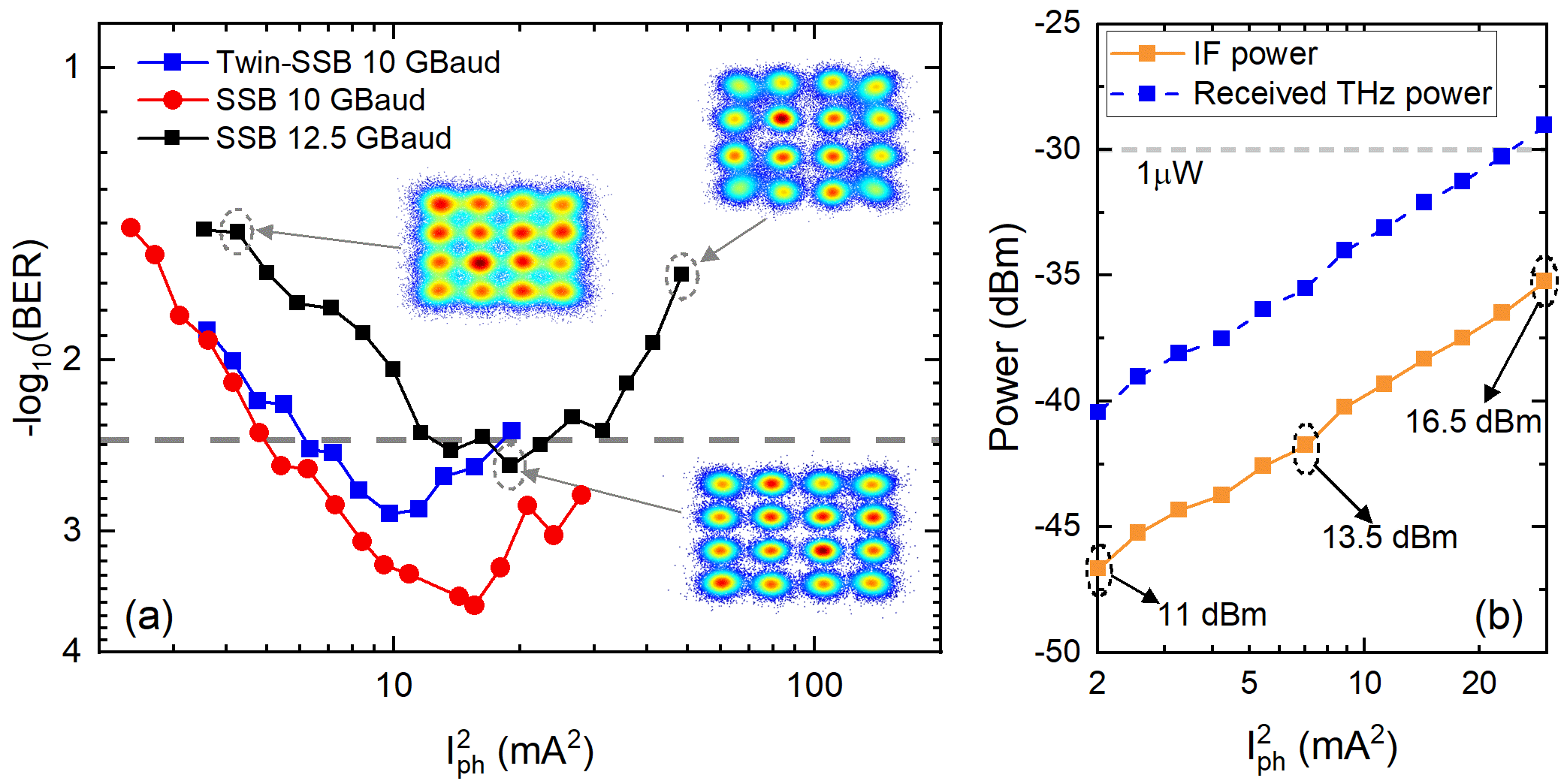}
\caption{(a) BER vs. squared photocurrent for each of the signals and data rates and (b) measured IF power after the enclosed receiver module and estimated received THz power vs. squared photocurrent. The insets in (a) show the constellation diagrams of the 12.5 GBd SSB signals for different values of photocurrent. The labels in (b) denote the input optical power to the UTC (the optical power was varied in 0.5 dB steps to increase the photocurrent).}
\label{50Gbit}
\end{figure}

Fig.~\ref{50Gbit} shows the BER curves obtained for each signal (i.e., twin-SSB and SSB) at 250 GHz. While both of them reached a BER below the FEC limit, the SSB signal achieved an error rate substantially lower than the twin-SSB technique. This is attributed to the higher peak-to-average power ratio (PAPR) of the latter and the limited dynamic range of our system (this will be discussed later). The PAPR of the twin-SSB signal (including carrier) was measured in Matlab to be 2.2 dB higher than that of the SSB signal. It is expected, hence, that using techniques to lower the PAPR of the twin-SSB signal will enhance the performance of this  scheme.

As a substantial margin over the FEC limit was obtained with the 10-GBd SSB signal, the symbol rate was subsequently increased to 12.5 GBd. The results of this transmission are shown in Fig.~\ref{50Gbit} with black squares. While a BER below the FEC was still recovered, the margin in this case was much lower than that for the 10-GBd signals. The main limitations preventing the transmission of higher-speed signals with the current system are two: (a) the available IF bandwidth at the Rx-RAU and (b) the limited output power and linear range of the UTC. The IF Rx-RAU bandwidth was limited due to a strong frequency response roll-off at high frequencies (shown in Fig.~\ref{IFandSNR} (a)) and a higher noise figure (NF) at low frequencies. The latter was measured indirectly by measuring the BER of signals down-converted to different IF frequencies. It was found that allocating the signal very close to DC severely degraded the quality of the signal. Also, the sub-channel from the twin-SSB signal which was down-converted closer to DC exhibited worse performance.   

The second cause can be seen from Fig.~\ref{50Gbit} (b), where the IF power (measured directly after the enclosed receiver module with no extra amplification) and the estimated received THz power (i.e., at the entrance of the module---calculated with the conversion loss data provided by Virginia Diodes) are plotted against the photocurrent squared. The maximum estimated received power is around -30 dBm. Furthermore, as can be seen from the constellations in Fig.~\ref{50Gbit} (a), saturation effects occurred for squared photocurrent values higher than 10 $\textrm{mA}^2$. 
These saturation effects originated in the UTC but it is not clear whether they are caused by thermal or space-charge effects \cite{Lin2019}. The
UTC was biased with a voltage of -2 V as lower voltages did
not improve the transmission performance.

By comparing Fig.~\ref{50Gbit} (a) and (b) (the latter was obtained by sending an unmodulated tone through the link) one can notice that the saturation effects were present only when transmitting data through the link. This suggests that the saturation point depends on the specific signal sent to the UTC, which agrees with what was observed in the transmissions of the SSB and twin-SSB signals. The presence of these saturation effects limited the available link budget and is the main reason why the experimental demonstration was realized over a short wireless link.

\textcolor{red}{With state-of-the-art UTCs and different types of antennas, the transmission distance reported here could easily be extended. In Table~\ref{tab:2} and~\ref{tab:3}, link budget calculations at a carrier frequency of 250 GHz are shown using two different approaches for the transmitter: (a) an array of antenna-integrated UTCs and (b) a high-gain antenna together with THz amplification. The same receiver (formed by a Cassegrain antenna, mixer and IF amplifier) is assumed in both cases.}

\begin{table}[t]
  \renewcommand{\arraystretch}{1} 
  \centering
  \caption{Achievable distance at 250 GHz and 50 Gbit/s with an integrated THz array antenna}
  \label{tab:2}
  \begin{tabular}
  {C{2.6cm} C{1.6cm} C{3.4cm}}
  \hline
  \rowcolor{mygray}
  Parameter & Value & Comments\\ 
  \hline
  \hline
  Antenna-integrated UTC & -10 dBm & \cite{Rouvalis2012} \\
  \hline
  Antenna array & +12 dB & $4 \times 1$ array  \cite{Sakano}\\ 
  \hline
  Transmission loss & 95.8 dB & \textbf{Distance = 5.9 m} (absorption negligible)\\ 
  \hline
  IF power & -53.8 dB & 55 dBi antenna \cite{Kallfass2015}; 10 dB conversion loss \cite{VirginiaDiodesa}; 5 dB implementation loss\\
  \hline
  IF input equivalent noise & -170 dBm/Hz & 5.4 dB NF\\
  \hline
  E$_\textrm{b}$/N$_\textrm{0}$ (at 50 Gbit/s) & 9.2 dB & = required E$_\textrm{b}$/N$_\textrm{0}$ for 16-QAM at BER = $3.8\times10^{-3}$\\
  \hline
  \end{tabular}
\end{table}

\begin{table}[t]
  \renewcommand{\arraystretch}{1} 
  \centering
  \caption{Achievable distance at 250 GHz and 50 Gbit/s with a Cassegrain antenna and THz amplification}
  \label{tab:3}
  \begin{tabular}
  {C{2.6cm} C{1.6cm} C{3.4cm}}
  \hline
  \rowcolor{mygray}
  Parameter & Value & Comments\\ 
  \hline
  \hline
  WR-3 waveguide UTC & -10 dBm & \cite{Song2012}\\
  \hline
  Amplifier & +15.3 dB & \cite{Griffith2017}\\
  \hline
  Cassegrain antenna & 55 dBi & \cite{Kallfass2015}\\
  \hline
  Transmission loss & 154 dB & \textbf{Distance = 1208 m} (absorption negligible; rain att.: 10 dB/km)\\ 
  \hline
  IF power & -33.7 dB & 55 dBi antenna \cite{Kallfass2015}; 10 dB conversion loss \cite{VirginiaDiodesa}; 5 dB implementation loss\\
  \hline
  IF input equivalent noise & -170 dBm/Hz & 5.4 dB NF\\
  \hline
  E$_\textrm{b}$/N$_\textrm{0}$ (at 50 Gbit/s) & 9.3 dB & = required E$_\textrm{b}$/N$_\textrm{0}$ for 16-QAM at BER = $3.8\times10^{-3}$\\
  \hline
  \end{tabular}
\end{table}

\textcolor{red}{With the $4\times1$ antenna array, a maximum coverage distance of almost 6 m is obtained (7 m if the 20\%-overhead soft decision (SD) FEC---BER threshold of $2\times10^{-2}$---is employed). This could be enough for indoor nomadic access in wireless local or personal access networks but falls short for a wireless bridge. However, as shown in Table~\ref{tab:3}, using state-of-the-art THz amplification and a Cassegrain antenna, a coverage distance of up to 1200 m (1350 m for the SD FEC) should be feasible even when assuming a heavy-rain scenario.}

\section{Conclusions and further work}
\label{sec.8}
The wide transmission bandwidths found at sub-THz frequencies make this part of the spectrum a strong candidate for ultra-fast wireless communications. Among the envisaged applications of sub-THz communications, wireless bridges have attracted significant interest due to the wide range of scenarios where they may be used. Due to their high integration potential with optical networks, THz wireless bridges based on photonic technologies offer a clear advantage over their electronic counterparts. In this paper, a photonic wireless bridge operating at 250 GHz and transmitting 20 Gbit/s is proposed and demonstrated. In order to mitigate the phase noise of the 4 free-running lasers present in such a link, the tone-assisted carrier recovery is used. Penalties of 0.15 and 0.46 dB with respect to the blind PNC algorithm are found when using this technique with aggregated Lorentzian linewidths of 28 kHz and 359 kHz, respectively, and 5 GBd 16-QAM signals. Based on these results and its simplicity, this algorithm may be a good  candidate for this type of link. The wireless bridge is also demonstrated in a WDM scenario, where 5 optical channels are generated and sent to the Tx RAU. In this unit, one channel is selected at a time and transmitted wirelessly. With this configuration, the full band from 224 GHz to 294 GHz is used. Finally, the results associated with the transmissions of 40 Gbit/s and 50 Gbit/s signals are presented. In this section, the main factors currently limiting the system capacity---UTC-PD output power and linear range and THz Rx IF bandwidth---are discussed.

In this demonstration, the data rate and transmission distance (10 cm) were mainly limited by the low power emitted by the UTC. Realistic distances ($>$ 100 m) may be achieved in future systems by improving the design of the UTC, integrating several of them in arrays, or using high-gain antennas together with THz amplification. Another point that needs to be addressed in photonic THz communications systems, is polarization stability. Options may be based on polarization control circuits or suitable coding techniques, but more research is needed on this topic. Another possibility is to use polarization multiplexing with polarization-diverse Tx and Rx RAUs. In this case, a twofold increase in the spectral efficiency will also be obtained, albeit with a significant increase in THz hardware. Another important parameter in these systems is frequency stability. While using free-running lasers may be attractive due to its simplicity and associated optical gain (assuming the optical LO is located at the Tx RAU), regulation on carrier frequency tolerance may prevent their use. In this regard, we will need to wait until specific regulations for the ultrawide bandwidth applications envisaged for sub-THz communications are developed. In terms of device technology, plasmonic modulators are promising candidates for THz/O conversion as they simplify the Rx RAU architecture and have very large modulation bandwidths. Future work could focus on comparing the performance of this type of modulators against approaches based on combining a Schottky barrier diode with a standard optical modulator.

\appendices

\section{Structure of the simulations}
\label{sec.9}
The same 5 GBd 16-QAM signals with $\alpha =0.1$ used in the experiments were used in the simulations. As phase noise was the only impairment of concern, only the PNC block---either that depicted in Fig.~\ref{DSP} (b) or the one described in \cite{Pfau2009a}---was used for signal demodulation. The penalty of the pilot tone-assisted technique for each $(\Delta\nu\times\textnormal{T})$ ratio was calculated 
as follows: first, the BER-verus-PTSPR-and-filter-bandwidth surface was obtained for different values of SNR. Second, after finding the lowest BER in each surface, a linear fitting of BER vs. SNR was performed. The SNR intersection of this fitting with a BER $=3.8\times10^{-3}$ was then obtained. Finally, the difference between this value and that given by equation (17) in \cite{Pfau2009a} was calculated to obtain the penalty shown in Fig.~\ref{FIRfilters} and~\ref{simulations}. in the case of the blind PNC technique, the same process but without the filter and PTSPR optimization step was carried out. A total number of 250000 bits were used for BER calculation. For the pilot tone-assisted technique symbols were gray-encoded, whereas differential encoding was applied to only the first two bits in the case of the blind PNC technique.

\section{Wireless transmission losses}
\label{sec.10}

\begin{figure}[t]
\centering
\includegraphics[width=8.8cm]{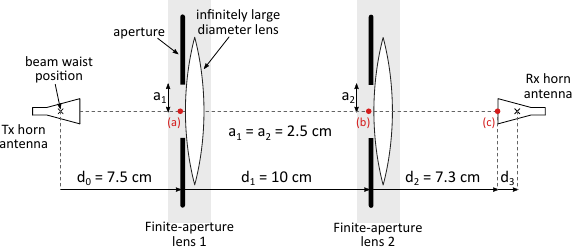}
\caption{Schematic diagram of the wireless transmission arrangement considered for path losses calculations.}
\label{Wsetup}
\end{figure}

\begin{figure}[t]
\centering
\includegraphics{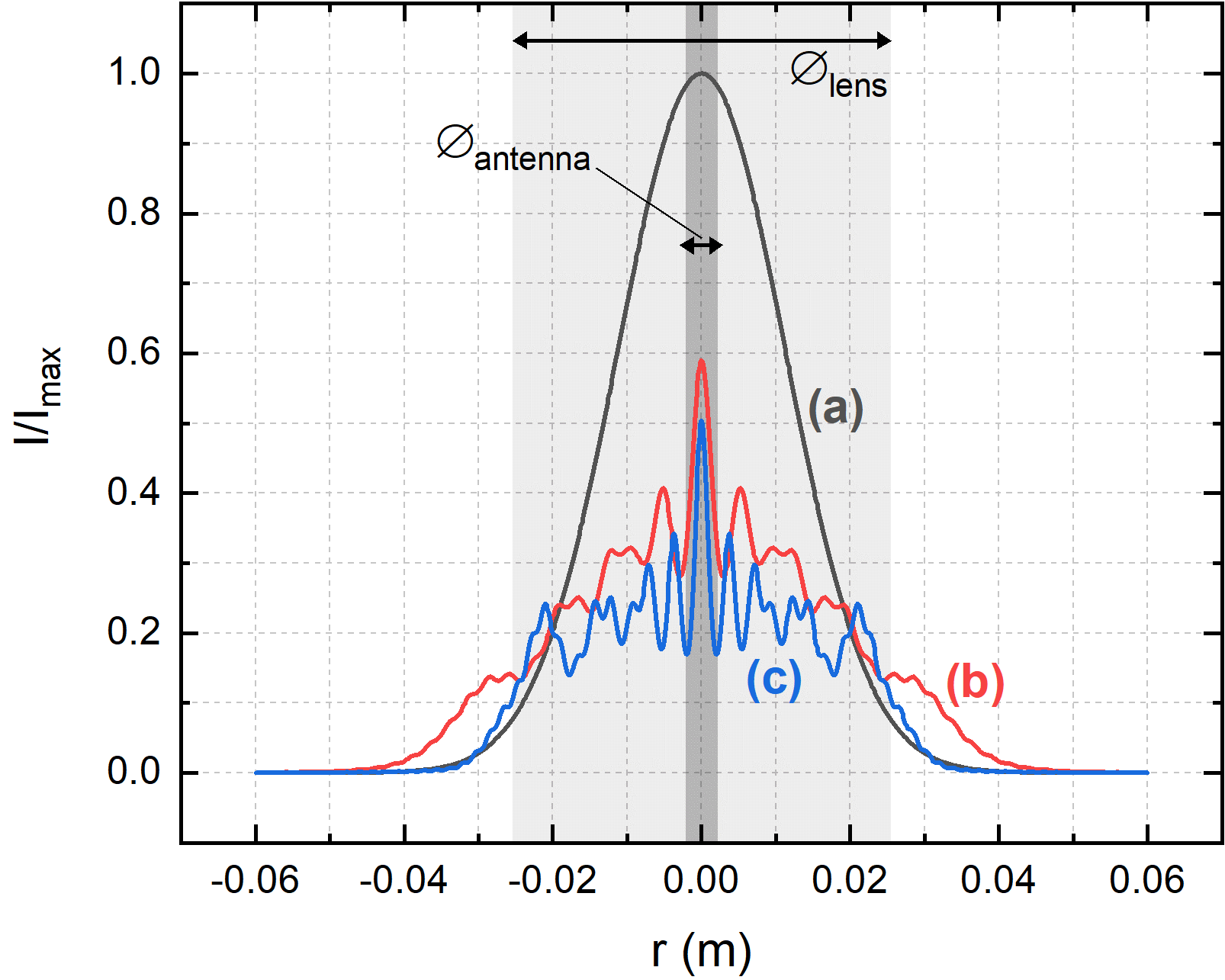}
\caption{Normalized radial intensity at the points marked in Fig.~\ref{Wsetup} with red points: (a) before the first lens, (b) before the second lens, and (c) at the input of the Rx horn antenna.}
\label{normintesity}
\end{figure}

\textcolor{red}{An attempt to calculate the loss of the wireless transmission system was made using the Huygens-Fresnel diffraction integral. The two finite-aperture lenses used in the experiment were modelled as apertures followed by infinitely-large diameter lenses as shown in Fig.~\ref{Wsetup}. The size and position of the imaginary beam waist inside the horn antennas was calculated with equations (1) and (2) in \cite{Nagatsuma2016f} and using values of \textit{L} (effective antenna length) and \textit{r} (aperture radius of the horn) of 11.3 and 2.2 mm, respectively. These values were provided by the manufacturer of the antennas. To reduce the long computation time associated with the numerical integration of the Fresnel integral, the method in \cite{Vicari1989}---which expresses the diffracted field as a linear combination of a set of values of the input field---was used. The normalized radial intensity at the points labelled in Fig.~\ref{Wsetup} is shown in Fig.~\ref{normintesity}. The diameters of the lenses and horn antennas used in the experiment are also shown for reference.}

\textcolor{red}{The coupling efficiency, $T$, between the incident field and the receiver antenna is calculated as:}

\begin{equation}
    T = \tau_x \tau_y = \tau_r^2
\end{equation}

\textcolor{red}{where $\tau_x$, $\tau_y$, and $\tau_r$  are the coupling efficiencies of the x, y, and radial directions, respectively. $\tau_r$ is given by \cite{Schwarz1984}:}

\begin{equation}
\label{eq:efficiency}
    \tau_r = \frac{\big|\int \textrm{E}_\textrm{1}(r)\textrm{E}_\textrm{2}^*(r) dr\big|^2} {\int\big|\textrm{E}_\textrm{1}(r)\big|^2 dr \int\big|\textrm{E}_\textrm{2}^*(r)\big|^2 dr},
\end{equation}

\textcolor{red}{where $\textrm{E}_\textrm{1}(r)$ and $\textrm{E}_\textrm{2}^*(r)$ are the wave functions of the incident beam and receiver horn antenna, respectively.} 

\textcolor{red}{A coupling efficiency of -20.6 dB was obtained using equation~\ref{eq:efficiency}. This value seems too low considering the much lower drop in SNR when going from point 2 to point 3 in Fig.~\ref{IFandSNR} (b). Assuming the FSPL is negligible when the two horn antennas are very close to each other (i.e., at point 2) and that receiver noise dominates over transmitter noise, the FSPL of the 0.1 m link should be equal to the SNR difference between these two points, which is approximately 1.5 dB. The large discrepancy between experiment and theory is attributed to the short diffraction distances and large apertures present in the transmission system, which might render the Fresnel approximation inaccurate. The Fresnel boundary, $R_n$, given by \cite{Selvan2017}:} 

\begin{equation}
  R_n = 0.62 \sqrt{\frac{D^2}{\lambda}},  
\end{equation}

\textcolor{red}{where $D$ is the largest dimension of the radiator (in this case the diameter of the lens), is equal to 0.2 m, which exceeds all propagation distances shown in Fig.~\ref{Wsetup}. The full evaluation of Kirchhoff's diffraction formula, therefore, would be more appropriate for this system, however this is beyond the scope of this work.}      

\ifCLASSOPTIONcaptionsoff
  \newpage
\fi


\bibliographystyle{IEEEtran}
\bibliography{references}

\end{document}